\documentclass{achemso}

\usepackage{amsfonts,amsmath,amssymb,mathbbol,mathtools}
\usepackage[english]{babel}
\usepackage{float}
\usepackage{graphicx}
\usepackage[hidelinks]{hyperref}
\usepackage{tikz}

\definecolor{lime}{HTML}{A6CE39}
\DeclareRobustCommand{\orcidicon}{
  \begin{tikzpicture}
    \draw[lime, fill=lime] (0,0) 
    circle [radius=0.14] 
    node[white] {{\fontfamily{qag}\selectfont \tiny ID}};
    \draw[white, fill=white] (-0.0625,0.095) 
    circle [radius=0.007];
  \end{tikzpicture}
  \hspace{-2mm}
}
\foreach \x in {A, ..., Z}{
  \expandafter\xdef\csname orcid\x\endcsname{\noexpand\href{https://orcid.org/\csname orcidauthor\x\endcsname}{\noexpand\orcidicon}}
}

\def\dehl{\ensuremath{{\Delta E}_\mathrm{HL}}}
\def\dethl{\ensuremath{{\Delta \widetilde{E}}_\mathrm{HL}}}

\def\ehs{\ensuremath{E_\mathrm{HS}}}
\def\els{\ensuremath{E_\mathrm{LS}}}

\title{Discovery of Spin-Crossover Materials with Equivariant Graph Neural Networks and Relevance-Based Classification}

\author{Angel Albavera-Mata\orcidA{}}
\affiliation{Department of Physics,
  University of Florida, Gainesville, Florida 32611}
\alsoaffiliation{Center for Molecular Magnetic Quantum Materials, Gainesville, Florida 32611}

\author{Pawan Prakash\orcidB{}}
\affiliation{Department of Physics,
  University of Florida, Gainesville, Florida 32611}
\alsoaffiliation{Center for Molecular Magnetic Quantum Materials, Gainesville, Florida 32611}

\author{Jason B. Gibson\orcidC{}}
\affiliation{Department of Materials Science and Engineering,
  University of Florida, Gainesville, Florida 32611}
\alsoaffiliation{Quantum Theory Project, University of Florida, Gainesville, Florida 32611}

\author{Eric Fonseca\orcidD{}}
\affiliation{Department of Materials Science and Engineering,
  University of Florida, Gainesville, Florida 32611}
\alsoaffiliation{Center for Molecular Magnetic Quantum Materials, Gainesville, Florida 32611}
  
\author{Sijin Ren\orcidE{}}
\affiliation{Department of Materials Science and Engineering,
  University of Florida, Gainesville, Florida 32611}
\alsoaffiliation{Center for Molecular Magnetic Quantum Materials, Gainesville, Florida 32611}
  
\author{Xiao-Guang Zhang\orcidF{}}
\affiliation{Department of Physics,
  University of Florida, Gainesville, Florida 32611}
\alsoaffiliation{Center for Molecular Magnetic Quantum Materials, Gainesville, Florida 32611}

\author{Hai-Ping Cheng\orcidG{}}
\affiliation{Department of Physics,
  Northeastern University, Boston, Massachusetts 02115}
\alsoaffiliation{Center for Molecular Magnetic Quantum Materials, Gainesville, Florida 32611}
  
\author{Michael Shatruk\orcidH{}}
\affiliation{Department of Chemistry and Biochemistry,
  Florida State University, Tallahassee, Florida 32306}
\alsoaffiliation{Center for Molecular Magnetic Quantum Materials, Gainesville, Florida 32611}
  
\author{S.B. Trickey\orcidI{}}
\email{trickey@ufl.edu}
\affiliation{Department of Physics and Department of Chemistry,
  University of Florida, Gainesville, Florida 32611}
\alsoaffiliation{Center for Molecular Magnetic Quantum Materials, Gainesville, Florida 32611}

\author{Richard G. Hennig\orcidJ{}}
\email{rhennig@ufl.edu}
\affiliation{Department of Materials Science and Engineering,
  University of Florida, Gainesville, Florida 32611}
\alsoaffiliation{Center for Molecular Magnetic Quantum Materials, Gainesville, Florida 32611}

\abbreviations{RMSE, MSE, MAE, R$^2$}


\begin{document}

\date{05 February 2025}
\newpage


\begin{tocentry}
  \includegraphics[width=\columnwidth]{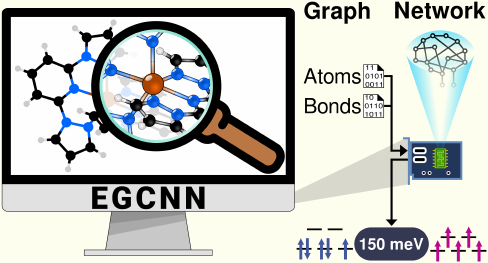}
\end{tocentry}


\begin{abstract}
Swift discovery of spin-crossover materials for their potential application in
quantum information devices requires techniques which enable efficient
identification of suitably spin switching candidates. To this end, we screened
the Cambridge Structural Database to develop a specialized database of 1,439
materials and computed spin-switching energies from density functional theory
for each material. The database was used to train an equivariant graph
convolutional neural network to predict the magnitude of the spin-conversion
energy. A test mean absolute error was 360~meV. For candidate identification,
we equipped the system with a relevance-based classifier. This approach leads
to a nearly four-fold improvement in identifying potential spin-crossover
systems of interest as compared to conventional high-throughput screening.
\end{abstract}

\maketitle

\section{Introduction \label{sec:introduction}}

First transition row $3d^4$ to $3d^7$ metal complexes may exhibit reversible
switching from a low-spin to a high-spin state. \cite{GG2004, H2004} Unlike
high-spin molecular magnets, for these types of complexes, two dominant effects
compete during the spin conversion, namely, the electronic occupation of the $d$
orbitals according to Hund's rule, and the filling of the t$_{2\mathrm{g}}$
lowest energy level. These mutually exclusive contributions result in two
possible ground states that depend directly on the strength of the ligand field.
More specifically, for strong ligand fields, the electrons preferably occupy
the t$_{2\mathrm{g}}$ orbitals, leading to a low-spin state. For the opposite
case that corresponds to a weak ligand field, if the strength of said ligand
field is larger than the electron pairing energy, then the electrons occupy the
maximum number of orbitals, according to Hund's rule, which results in a
high-spin state. Interestingly, whenever the strength of the ligand field and
the electron pairing energy share nearly the same order of magnitude, a small
external perturbation can overcome either effect and switch a metal complex to
a low- or high-spin state. \cite{NB2016} The choice of ligands The attractive
feature of such reversible switching is its potential use in display devices,
mechanical actuators, high-density memory storage, optoelectronics, sensing
devices, or spintronics. \cite{KKJ1992, KM1998, GKG2002, MVL2003, LGG2004,
LSGTMMB2008, S2011, QFSSMSNB2014, R2014, JLRRPDWG2015, MRSMQNSB2016,
MMSRRSBCMCLNNSMB2018, CP2021}

From a computational perspective, these materials pose a challenge even to
high-level wave function theories. \cite{DARH2012, DAS2013, SKSBHB2018,
PFHP2018, FGRTN2020, ZT2020, DMP2021, DMP2022} There are many considerations,
one of which is that the accuracy scale of these high-level theories is
comparable to the typical energy difference between the two spin states of
approximately 10~kJ\,mol$^{-1}$, or about 100~meV. This difference, in turn, is
close to the accuracy scale of the electronic structure method of choice.
\cite{K2013, K2019}

From the synthesis perspective, the vast combinatorial chemical space engendered
by the metallic core, ligands, coordination number, functional groups, size of
the complex, strength of the intra- and intermolecular interactions, etc., poses
a major challenge to efficient large-scale screening for the selection of promising
materials suitable for specific applications. \cite{AHBMTCPHDG2010,
MMRFNB2019, BB2020, MLIY2021}

Development of machine learning strategies for elucidating the intricacies of
the relationships between this vast material space and the spin splitting energy
is an active area of research. \cite{JK2017, JCK2018, TYLNJDK2020} Ref.
\citenum{NDTLSK2021} highlights the complexity of such endeavors. Additionally,
because of limited reports of experimental spin-crossover pursuits that ended up
not finding spin switching materials, it is arguable that the diversity of
materials and availability of experimental data limits the extent to which general
trends can be identified and applied. \cite{BL2003, KL2010} As a result, a
compromise takes precedence, namely, to restrict data to experimental
observations \cite{VKTDK2024} or use results from high-throughput computations.
\cite{TYLNJDK2020, KRTBMSKV2024} Refined learning from experiments may be achieved
by considering only specific families of materials of interest, at the cost of
losing transferability to complexes of varying nature, and of not being able to
account for unsuccessful candidate materials. Conversely, high-throughput
calculations of the spin-crossover energy with, for example, Kohn-Sham density
functional theory, are feasible. No general protocol exists, however, to select
an adequate exchange-correlation density functional approximation, a choice
crucial to the accuracy of the results. \cite{PT2004, PSW2013, DRM2017, CVR2018,
CR2019, RHBM2020}

Database screening is among the more popular approaches for culling material
candidates. With potentially numerous samples, each of which involves large
complicated molecules, machine learning techniques such as decision trees,
kernel-based algorithms, or artificial neural networks, can aid in uncovering
structure-property relationships, thereby accelerating the identification of
candidates with promising physical chemical properties. \cite{ZG2021, MJOKEB2021,
DNK2022, CSH2022} Among these choices, graph neural networks provide a feasible
architecture for encoding relevant physical and chemical descriptors for
molecule-based and atom-based applications. \cite{KW2016, BBLSV2017, YLLS2023,
KPKT2024} These advantages make graph models of particular interest for the
description of spin-crossover materials.

This work thus is focused on assessing the performance of a specific type of
graph model that is capable of learning the importance of symmetry
representations, \cite{MBSL2018, TSKYLKR2018, KKN2019} namely, an equivariant
graph convolutional neural network. \cite{SHW2021, SHW2021Github} The task is
the prediction of the sometimes elusive energy difference between the accessible
spin states for a modest set of transition metal complexes. We show that such a
network, with only 915 trainable parameters, can learn the relationship between
the structure and spin-state switching behavior for our data set. We provide
evidence that the model achieves this ability by learning the importance of the
coordination shells surrounding the metallic core. We conclude that the network
may provide an efficient alternative for candidate screening, and demonstrate
its usefulness on a larger data set as proof of concept.

\section{Methodology \label{sec:methodology}}

We first describe the selection criteria for the material constituents of the
dataset obtained from the Cambridge Structural Database. \cite{GBIL2016} The
screening strategy was implemented by identifying molecular structures
containing Cr, Mn, Fe, or Co metal center and excluding all materials with more
than fifty non-hydrogen atoms in the asymmetric unit. Then the fundamental
properties of the resulting 1,439 materials were computed via Kohn-Sham density
functional calculations that used the r$^2$SCAN exchange-correlation functional
approximation \cite{FKNPS2020} since it has been shown to provide
reasonable energy differences for metal complexes. \cite{Mejia_Rodriguez_2020}
The electronic structure calculations used \textsc{Vasp} 6.2, \cite{KF1996} with
the set of projector augmented wave potentials for the outermost electrons as
valence states, except for the transition metals, which were treated with the
potentials that include the $4s$ semi-core states as valence states. We set the
plane wave kinetic energy cutoff to 520~eV and used an auxiliary support grid
for the evaluation of the augmentation charges. The precision tolerance labeled
as accurate was selected, non-spherical corrections were activated, and $10^{-6}$
and $10^{-3}$ eV were used as thresholds for the electronic steps and the
Gaussian smearing width, respectively. All geometries were optimized with the
conjugate gradient algorithm until force magnitudes were smaller than $10^{-2}$
eV\,{\AA}$^{-1}$. The rest of the computational parameters were left at default
values.

\begin{figure}
  \includegraphics[width=0.5\columnwidth]{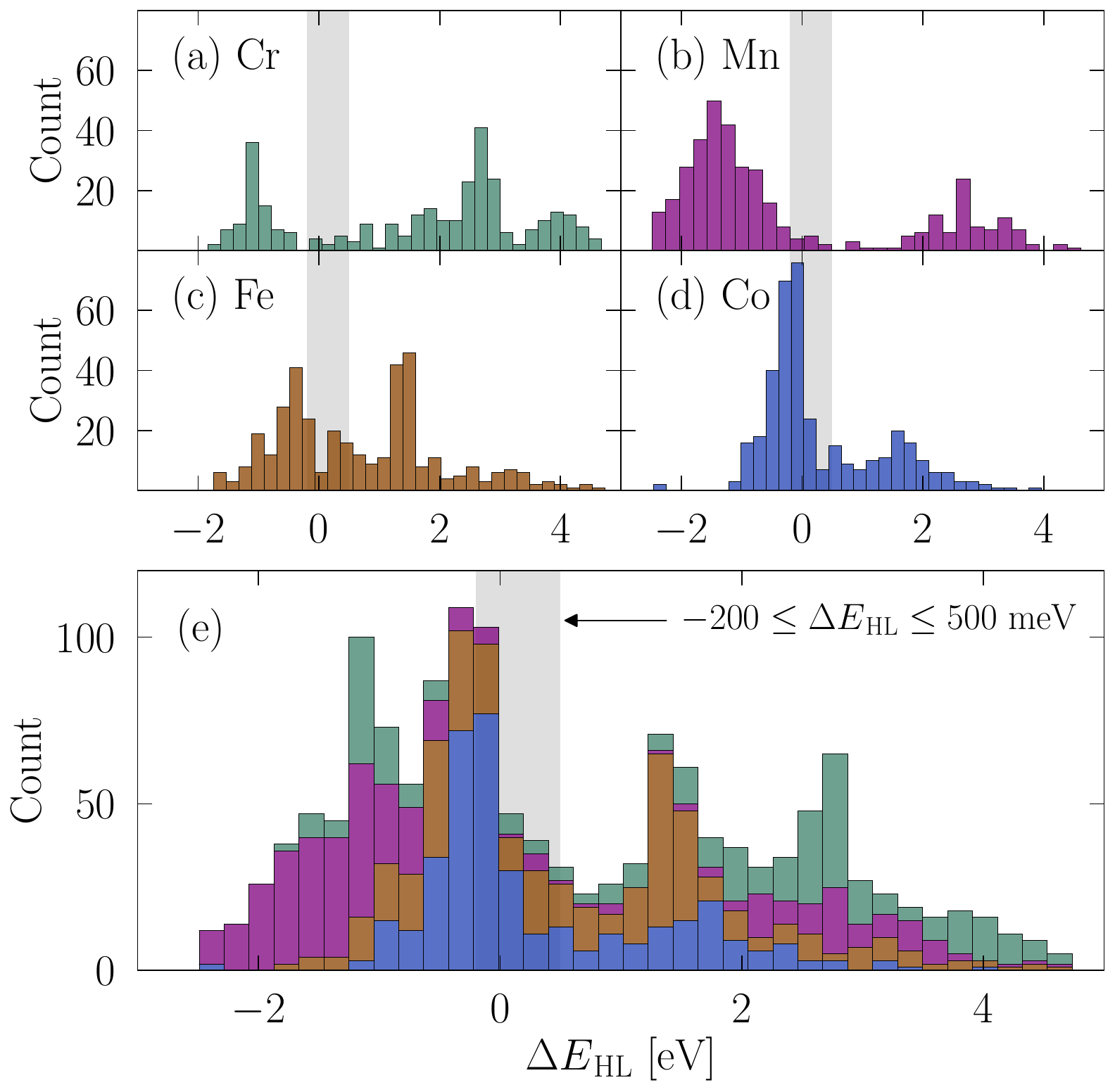}
  \caption{Distribution of the calculated crystalline spin-crossover energies
           per molecule for the 1,439 species in the data set. Counts for each
           interval are depicted for (a) 316 Cr-, (b) 374 Mn-, (c) 372 Fe-, and
           (d) 377 Co-containing complexes, and (e) the dataset as a whole. The
           shaded area illustrates the choice of region of interest for
           spin-state switching candidates.
  \label{fig:1}}
\end{figure}
It is essential that the spin transition in our study be characterized uniquely
for the structure relaxations and the crossover energy of a candidate system of
our interest. We required that the relevant spin-switching complexes satisfy
three conditions, namely that,
\begin{enumerate}
  \item the system ground state is low-spin,
  \item the spin and energy ordering of the states are consistent, i.e. the
        first excited state is high-spin, and
  \item the maximum low-spin to high-spin conversion energy per molecule is
        bounded to 500 meV, or $\approx$ 4000 cm$^{-1}$.
\end{enumerate}
Criterion 3 is generous, considering the error magnitudes commonly associated
with electronic structure calculations for these materials.\cite{MAFCCHT2022}
The resulting distribution of the crystalline spin transition energies is
depicted in Figure \ref{fig:1} for 316, 374, 372, and 377 complexes with a Cr,
Mn, Fe, and Co center, respectively. The data set has a mean spin-state
switching energy of 538.5~meV and a standard deviation of 1.68~eV. The full list
of entries is included as Supplementary Material. The aforementioned energy
differences are for crystalline computations, but to keep with standard
spin-crossover analysis, for the eventual learning step, we isolated the
molecular unit for each material based on the reported experimental coordinates,
and assigned $-200 \leq \dehl \leq 500 $ meV as the range of interest for
spin-state switching candidates. Here, {\dehl} is the spin-crossover energy for
the metal complexes with total spin $S$ increments given by
\begin{equation} \label{eq:1}
  \mathrm{S} = \begin{cases}
                 1 \rightarrow 2 & \text{for $3d^4$ ions (Cr$^{2+}$ and Mn$^{3+}$}),\\
                 \frac{1}{2} \rightarrow \frac{5}{2} & \text{for $3d^5$ ions (Mn$^{2+}$ and Fe$^{3+}$}),\\
                 0 \rightarrow 2 & \text{for $3d^6$ ions (Fe$^{2+}$ and Co$^{3+}$}),\\ 
                 \frac{1}{2} \rightarrow \frac{3}{2} & \text{for $3d^7$ ion (Co$^{2+}$} \, ).
               \end{cases}
\end{equation}
The transition profile therefore can be represented as the total energy
difference $\dehl = \ehs - \els$, where {\els} and {\ehs} are the total energy
per molecule for the low-spin and high-spin states, respectively.

Regarding the choices for the machine learning model, we used the equivariant
graph neural network as designed and implemented by Satorras, Hoogeboomand, and
Welling, \cite{SHW2021, SHW2021Github} with a modified attention layer, $\phi_a$,
described below. This scheme considers the system of interest to be represented
as a graph $\mathcal{G} = \mathcal{V} + \mathcal{E}$, with the set of nodes
${\upsilon}_{i} \in \mathcal{V}$ and the set of edges
${\varepsilon}_{i,j} \in \mathcal{E}$. We follow those authors' notation for
clarity. Specifically for our purposes, for each system with $N$ atoms, the
feature node embeddings $\mathbf{h} \in \mathbb{R}^{N \times 1}$ list the atomic
number $z$, whereas $\mathbf{x} \in \mathbb{R}^{N \times 3}$ correspond to the
3D Cartesian coordinates. Both $\mathbf{h}$ and $\mathbf{x}$ are associated with
each of the graph nodes. We emphasize that, by design, this architecture
preserves equivariance to translations and rotations on $\mathbf{x}$, and
equivariance to permutations on the set $\mathcal{V}$. See Appendix A of Ref.
\citenum{SHW2021} for proof.

To distinguish differences across distinct chemical functional groups, we also
included the set of node attributes $\mathbf{v} \in \mathbb{R}^{\mathcal{V}
\times 1}$ and edge attributes $\mathbf{e} \in \mathbb{R}^{\mathcal{E} \times 1}$,
that list the oxidation state for each node and bond order for each edge,
respectively. \cite{IUPAC} In particular, the oxidation state for each atom was
determined by pair-wise comparison of the electronegativity between its
heteronuclear bonds, whereas  the bond order is the sum of the products of the
corresponding atomic orbital coefficients over all the occupied bands, rounded to
its closest integer. The data set with these descriptors is available from the
authors upon request. We restricted the model to a single convolutional layer to
prevent overfitting due to the limited size of our data set. As a result, the
model has only 915 learnable parameters. Furthermore, we used the non-linear
activation function known as \textit{tanhshrink}, and trained the network for a
maximum of $10^{4}$ epochs with an early stop during validation. \cite{P1998}
The parameters were optimized with the adaptive moment estimation in
stochastic gradient descent, \cite{KB2017} with a weight decay of $10^{-6}$,
considering an initial learning rate of $10^{-2}$, and reduced on plateaus
during training using a threshold of 1~meV. \cite{KSH2017} Lastly, the edge
attention $\alpha_{i,j}$ is defined as
\begin{equation} \label{eq:2}
  \alpha_{i,j} = \phi_a(\mathbf{h}_i, \mathbf{h}_j \vert\vert \mathbf{e}_{i,j})\, ,
\end{equation}
where $\phi_a$ is a weight matrix applied to every edge for the pair of nodes 
$i,j$, and $\vert\vert$ denotes the concatenation operation between
$\mathbf{h}_i, \mathbf{h}_j$ and $\mathbf{e}_{i,j}$.

To establish a baseline model, we chose the gradient boosting decision tree from
the scikit-learn library.\cite{sklearn} The optimized hyper-parameters using a
grid search resulted in a learning rate of $7 \times 10^{-2}$ using the Huber
loss for a total number of 350 estimators. The tree consists of a maximum depth
of 64 estimators for a maximum splitting feature equal to the squared-root of
the number of features. We also used a minimum of four samples at a leaf node
for a minimum of 52 samples required to split an internal node. The model is
available from the authors upon request.

For both the graph network and baseline model, the whole data set was divided
into 90\,\% training fraction and 10\,\% holdout fraction for testing. The
90\,\% training set was used to optimize the hyperparameters using grid search
in a ten-fold cross-validation without overlapping samples to ensure that each
element is tested once. The training and testing splittings were identical for
the two models. The behavior of the loss as a function of epochs for the neural
network is depicted in Figure \ref{fig:2}(a). The training loss clearly
decreases as the number of epochs increases. The testing loss, on the other hand,
decreases during the first 70 epochs, then increases slightly through the
following 130 epochs, followed by a second decrease as the number of epochs
increases. After nearly 2,000 epochs, the testing loss increases marginally and
plateaus, showing early signs of overfitting. It is important to note that this
is a retrospective analysis, because, as stated previously, we did not implement
an early stop for the final model. However, considering that the mean test loss
increase with respect to epoch 2,000 is on the order of $10^{-2}$ meV, the
over-fit can be safely disregarded.

\section{Results and discussion \label{sec:results}}

\begin{figure}
  \includegraphics[width=0.5\columnwidth]{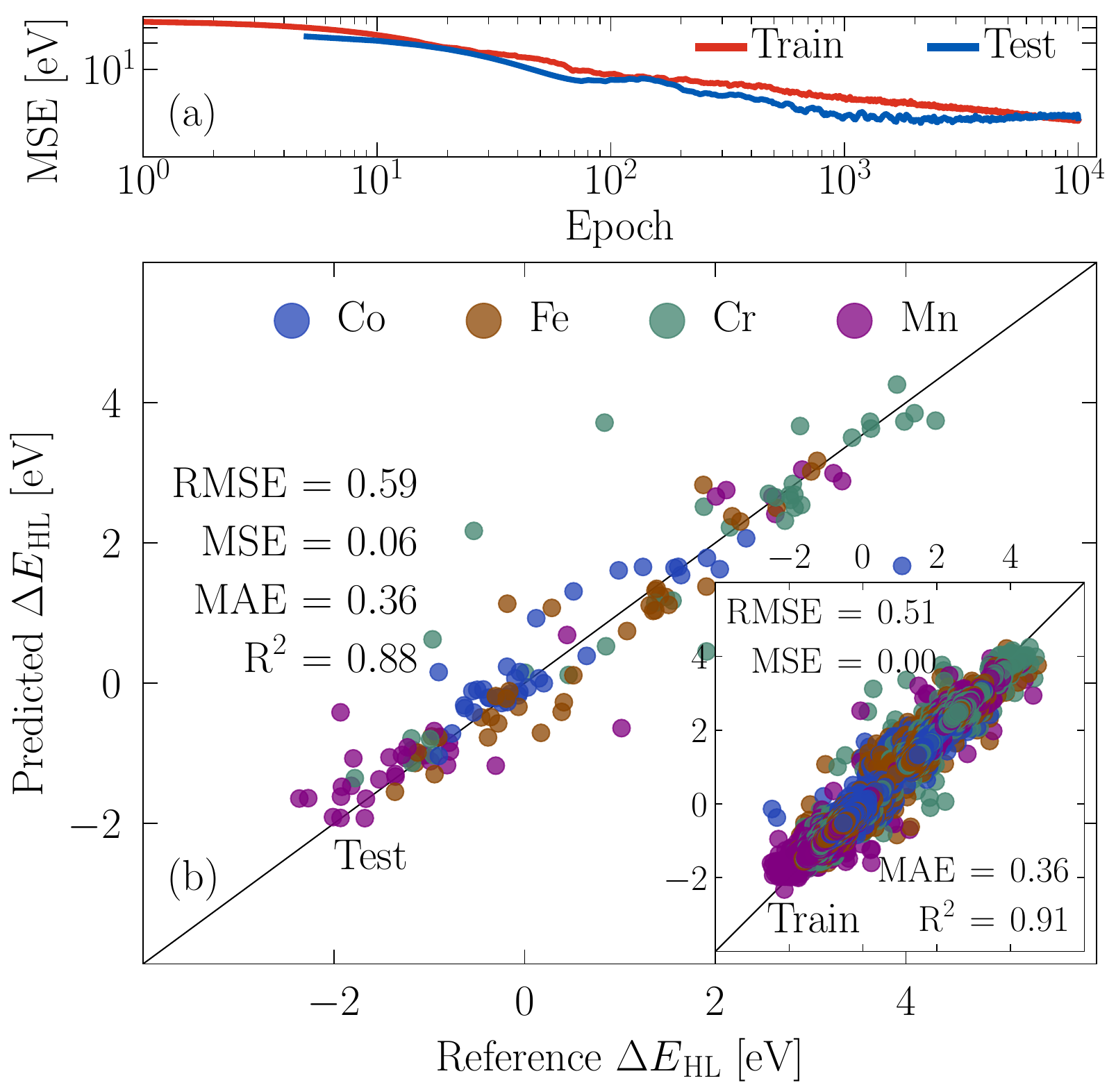}
  \caption{(a) Behavior of the mean signed error loss with a summation reduction
            as a function of epochs for the training and testing sets. (b)
            Correlation between the reference and predicted values using the
            equivariant graph network for both the testing and training sets. The
            measures of central tendency are the root mean squared error (RMSE),
            mean signed error (MSE), and mean absolute error (MAE). R$^2$ is the
            coefficient of determination.
  \label{fig:2}}
\end{figure}
We begin with evaluation of the predictive capability of the graph model.
Comparisons for the agreement between the reference and predicted {\dehl} values
are reported in Figure \ref{fig:2}(b) for the testing set. The training sample
comparison is in the inset. For statistical purposes, we used the interquartile
range between the first and third quartile to characterize the spread and
variability of the predicted {\dehl} for identification of outliers. The
Cr-containing complexes show a total of four outliers, namely, LORBUQ, CYCPCR,
ALUVIN, and CTNSCR; three for the Mn molecules, CAXTIH, UDOKON, and VEKFIA;
three for Fe complexes, FIXSAH, HEYNEE, and CITCUD; and two for Co molecules,
WIYJEV, and SETJEI. This distinction can be partially attributed to the abundance
of air-stable complexes containing the Cr$^{3+}$ ion ($3d^3$), which does not
exhibit spin crossover, as compared to much rarer and more sensitive complexes
of the Cr$^{2+}$ ion ($3d^4$) that can undergo spin crossover. Most of molecules
containing Mn, Fe, and Co correspond to octahedral complexes and metallocenes.
Putting that aside, although Figure \ref{fig:2}(a) shows early signs of
overfitting, we see reasonable agreement between the testing and training
predictions reported in Figure \ref{fig:2}(b), indicating that the effects are
minimal.

\begin{figure}
  \includegraphics[width=0.5\columnwidth]{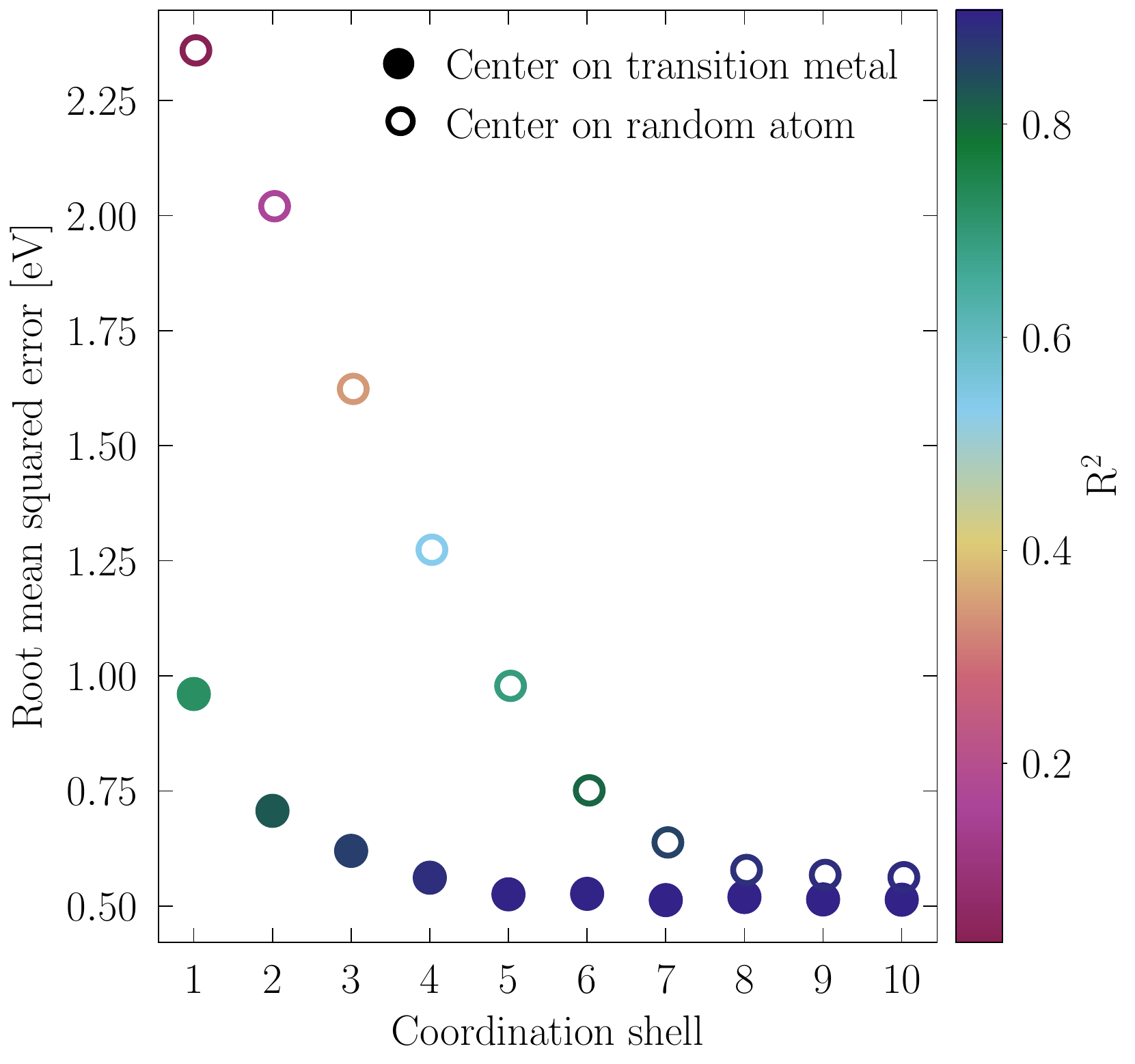}
  \caption{Behavior of the RMSE and R$^2$ as a function of graphs built with
           nodes and edges for sequentially increased numbers of coordination
           shell approximations. Two reference types were chosen, namely,
           coordination shells starting at the metallic center or at a random
           atom of the molecule. For both cases, the reference atom was kept
           fixed once picked.
  \label{fig:3}}
\end{figure}
From the implementation perspective, our goals include learning chemically
relevant features and understanding how the neural network makes the predictions.
We expect the modified attention layer in Eq. \eqref{eq:2} to be able to provide
the means for distinguishing molecular sites that are more strongly associated
with the spin transition from other sites.

With that in mind, we sampled a series of progressively more distant coordination
shells surrounding the transition metal center, under the consideration that the
largest number admissible of coordination shells must not exceed the number of
neighboring atoms in a complex. In each case, sub-graphs were built that
included the nodes and edges of all the molecules in that set of shells. Figure
\ref{fig:3} shows that both the RMSE and $R^2$ converge rapidly as the number of
coordination shells increases. Results remain nearly unchanged after five shells.

A distinctly different trend is observed if a random atom is used as the
coordination shell center. For that choice, Figure \ref{fig:3} shows that the
mean errors roughly double, as does the number of coordination shells needed to
achieve the same quality results as in the metal-centered case. The $R^2$ values
also deteriorate. It must be noted, however, that since the data set includes
medium-sized molecules, using ten coordination shells essentially covers the
whole molecule for most of the samples. The contrasting tendencies for the two
approaches nonetheless show that the model learned the local importance of the
transition metal core dominance for the spin transition.

\begin{figure}
  \includegraphics[width=0.5\columnwidth]{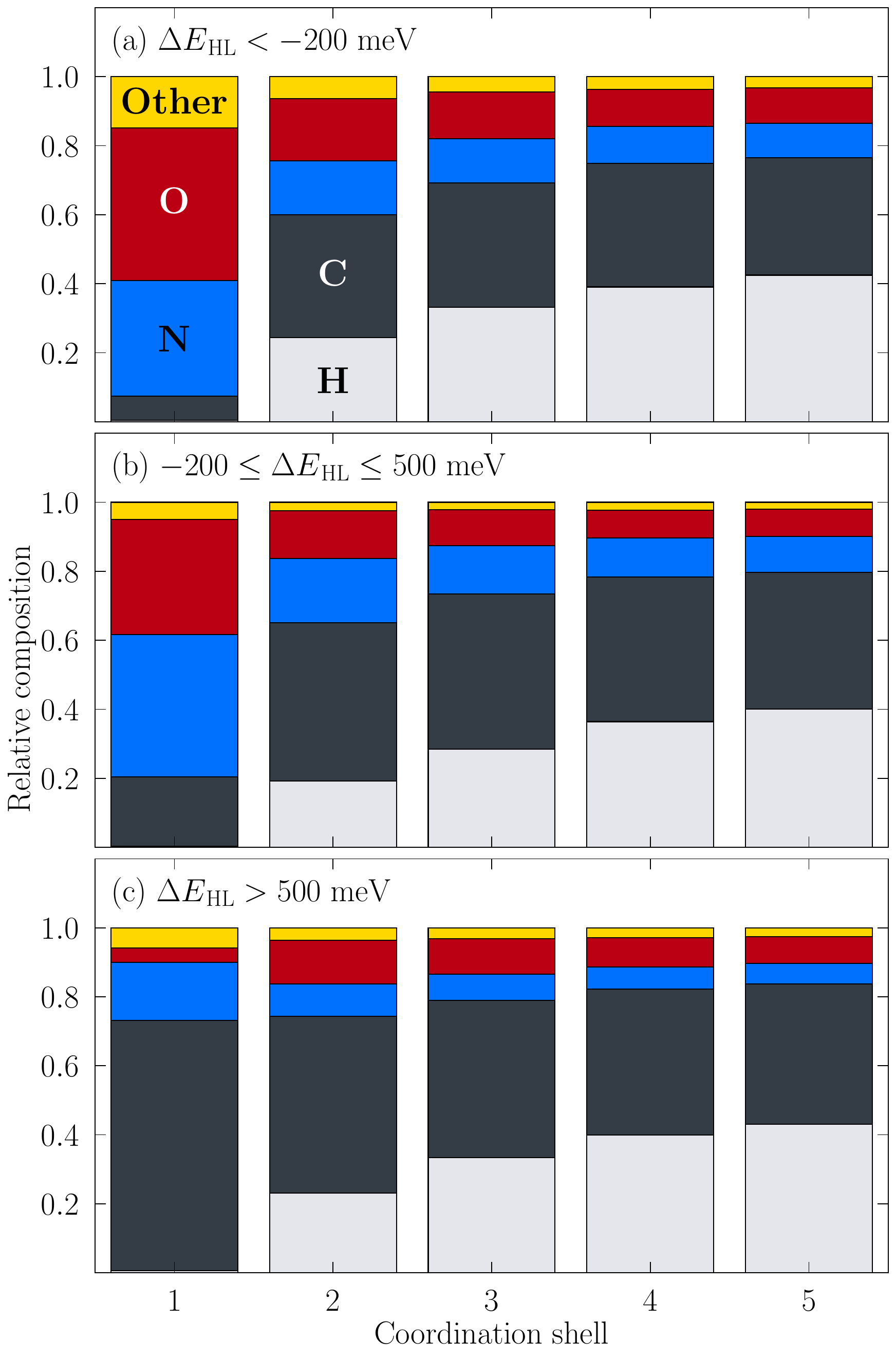}
  \caption{Normalized compositions for Carbon, Nitrogen, Oxygen, Hydrogen and
           other atoms, such as B, F, P, S, Cl, I, etc., as a function of the
           number of coordination shells centered on the transition metal for
           the species showing (a) high-spin states with $\dehl < -200$ meV, (b)
           spin-switching candidates, with $-200 \leq \dehl \leq 500$ meV, and
           (c) low-spin states $\dehl > 500$ meV.
  \label{fig:4}}
\end{figure}

In addition, examining the characteristics of the element composition of the
complex core may provide insights about the chemical features that the model
learned to determine {\dehl}. For that, we focused attention on the first five
coordination shells that show a more rapidly varying RMSE in Figure \ref{fig:3}.
For each progressively larger shell, we counted the total number of distinct
elements and computed their cumulative summation, we then calculated the relative
composition per element by dividing the element count by the total number of
atoms in that shell. Figure \ref{fig:4} shows the results for such analysis, with
particular emphasis on the Hydrogen, Carbon, Nitrogen and Oxygen atoms that
represent the largest compositions across all coordination shells.

For the purposes of the following discussion, we will omit the Hydrogen atoms
because their relative composition increases rapidly with the number of
coordination shells, and because they do not provide meaningful information
regarding ligand composition. The element decomposition for the species
identified as high-spin states is shown in Figure \ref{fig:4}(a). Here we observe
that the O atom is most predominant, followed by N, and a noticeable presence of
other atoms like B, F, P, S, Cl, or I. This composition agrees with the
spectrochemical series, \cite{T1938i, T1938ii, TK1938iii, HT1938iv, YT1938v}
where the weak field of ions such as O$_{2}^{2-}$, I$^{-}$, Br$^{-}$, Cl$^{-}$,
NO$_{3}^{-}$, or F$^{-}$ results in high-spin states.

For the species identified as spin crossover candidates depicted in Figure
\ref{fig:4}(b), on the other hand, we see nearly three times fewer weak-field
atoms, and a large composition of N and O atoms where N is the slightly
predominant element, as opposed to the fewer population of C atoms that increases
for larger shell counts. These findings agree with the element composition of the
typical ligands used to synthesize these metal complexes, \cite{NSMO2007, GS2014,
SPCS2015, HHP2016, FVCMPCCLRJ2017, HSB2018, WLIY2021} whereas the C atoms that
appear in the first coordination shell can be attributed to the presence of
metallocenes.

Regarding the low-spin species, Figure \ref{fig:4}(c) shows that the C atoms
account for roughly 75\% of the total composition for the first coordination
shell, followed by N, and a nearly equal composition of O and other atoms that,
however, becomes more populated by O atoms for larger coordination shells. Once
again, these findings agree with the spectrochemical series where the strong
field of ions alike CH$_{3}$CN, NH$_{3}$, NO$_{2}^{-}$, CN$^{-}$, and CO produce
low-spin states.

The trends observed for N and O, for both spin-switching and not spin-switching
complexes depicted in Figure \ref{fig:4}, remain nearly unchanged for a larger
number of coordination shells, whereas the number of C atoms increases rapidly
as we sample a larger subspace of the ligand structure.

The motivation for using coordination shells is to provide qualitative insight
into the effects of ligand field stabilization. At the same time, the spin
transitions of interest are modulated by a delicate balance between the metal
ion properties and the strength of the ligand field, \cite{STHN1972, SS1974,
KSK1986, PDWTT2001, T2001, SNM2006, K2013, PSW2013, MK2015, KKMDBLDH2016} in
which vibrational contributions to the free energy must be taken into account
for detailed representation of thermochemical behavior. Those considerations
are well beyond the scope of the present work. Here we focus solely on using the
energy difference for elucidating tendencies.

\begin{figure}
  \includegraphics[width=0.5\columnwidth]{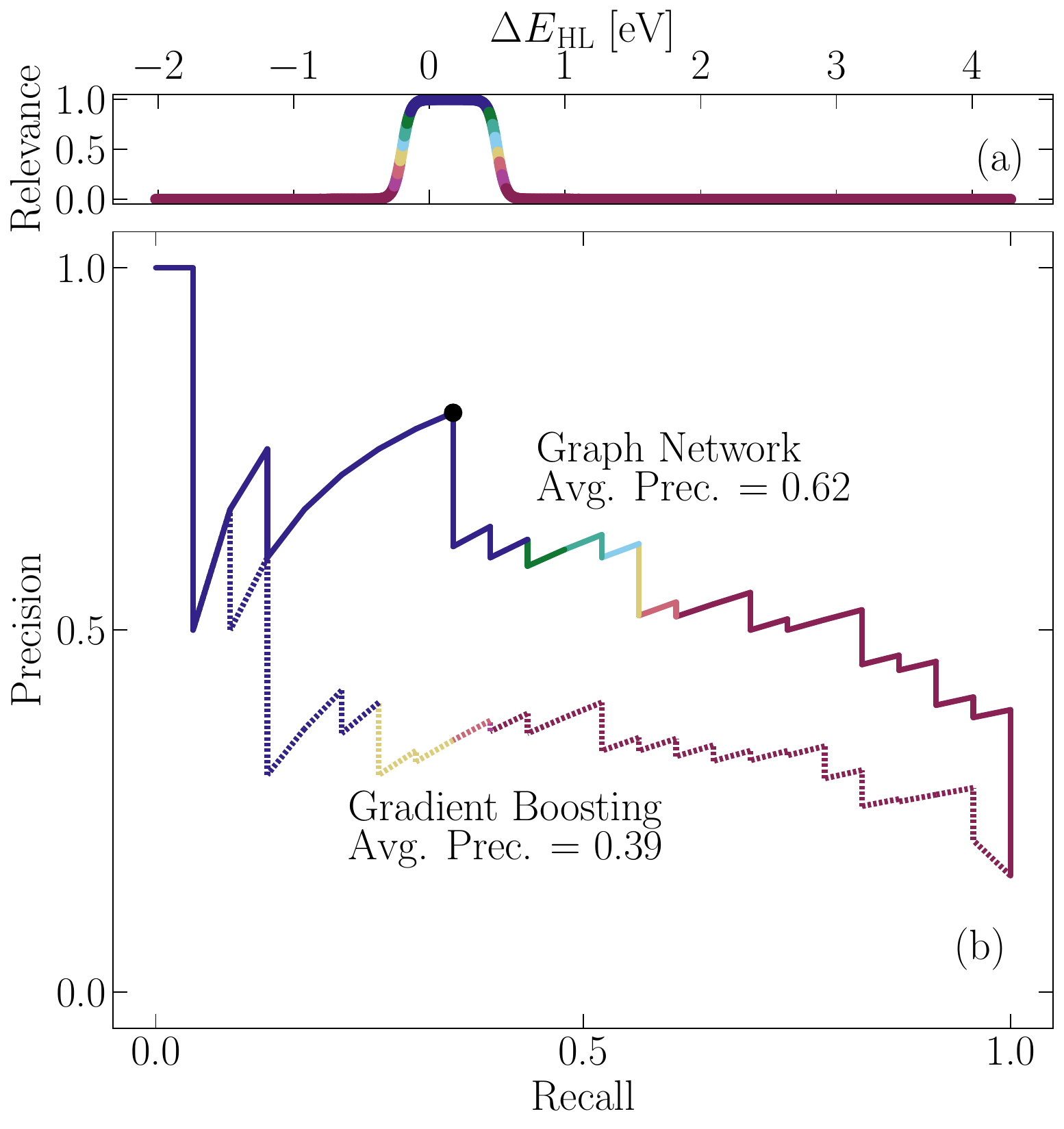}
  \caption{Classification of molecular metal complexes as potential candidates
           for spin crossover (a) based on the energy difference between the
           high-spin and low-spin states, and (b) the precision-recall curve for
           the testing set as a function of the choice of relevance for the graph
           neural network and the gradient boosting base-line method. Identical
           data splittings were used for both methodologies.
  \label{fig:5}}
\end{figure}
Because the motivation for constructing and training the network is its
subsequent use to search for potential spin-crossover complexes, we need to
convert the regression into a classification model. Therefore, establishing the
precision-recall trade-off becomes relevant. The precision measures the
correctness of the model for actual spin-crossover predictions, whereas the
recall is associated with how many relevant complexes the model recovers. We used
the method proposed by Torgo and Ribeiro \cite{TR2009} to turn the continuous
regression into a classification process. To do so, we must define the importance
for the target class, referred to as relevance, through the use of a continuous
relevance function scale that maps the original continuous domain to the discrete
target class. Recall that the energy interval of interest, defined generously, 
is $-200 \leq \dehl \leq 500$ meV. A basic approach is to define the relevance
$f(Y)$ for a given {\dehl} by means of two sigmoid functions, one for the lower
and another for the upper limit, expressed by
\begin{equation} \label{eq:3}
  f(Y) = \frac{1}{1 + \exp\left[{-s\cdot(Y - c)}\right]}\,,
\end{equation}
where $c$ is the center of the sigmoid and defines the threshold where larger
values of the target variable $Y$ start to become more relevant. By construction,
both thresholds have a relevance equal to 1/2, as depicted in Figure
\ref{fig:5}(a). The remaining variable $s$ is related to the slope of the sigmoid
functions. It is fixed by the energy difference resolution, which we chose at
1~meV. Figure \ref{fig:5}(a) shows clearly how the relevance is larger for energy
differences in between the interval of interest, and tends to zero otherwise.

Having stipulated the mapping between {\dehl} and its relevance, the
precision-recall curve, depicted in Figure \ref{fig:5}(b), is generated readily
by determining the combinations between the true and false classes with the
positive and negative outcomes. Once more, we used the testing set for our
analysis. Since we are interested in minimizing the the number of redundant
calculations, the optimum value for the relevance $f(Y)$ threshold is such that
it maximizes the precision while retaining a reasonable recall. This choice is
depicted with the dot in Figure \ref{fig:5}(b), which corresponds to a precision
of nearly 80\,\% and a recall of roughly 35\,\% for a relevance of roughly 0.90.
That, in turn, accounts for a difference of 68~meV  with respect to the original
lower and upper limits, resulting in the tighter $-132 \leq \dehl \leq 432$ meV
interval and a false positive rate of just 1.7\%.

In addition, we can make use of Bayes' theorem to determine the conditional
probability $P(\dethl \vert \dehl)$ for finding a candidate with predicted
{\dethl} that truly is a crossover complex with a calculated {\dehl}. This is
achieved through the expression $P(\dethl \vert \dehl) = \allowbreak P(\dethl)\,
P(\dehl \vert \dethl)/P(\dehl)$, where $P(\dethl) = 167/1296$ is known as the
prior probability that, in our case, reduces to the relative population of
spin-crossover complexes in the set, whereas $P(\dehl \vert \dethl) = 81/167$ is
the likelihood probability for observing spin-crossover complexes that also are
identified as switching complexes in the reference data. Lastly, $P(\dehl) =
P(\dehl \vert \dethl)\,P(\dethl) + P(\dehl \vert
{\Delta \tilde{E}^c}_\mathrm{HL})[1 - P(\dethl)]$ is the evidence probability for
observing a spin-crossover candidate expressed as a function of the complement of
{\dethl}, ${\Delta \tilde{E}^c}_\mathrm{HL}$, where $P(\dehl \vert
{\Delta \tilde{E}^c}_\mathrm{HL}) = 1 - 1065/1129$. Substitution of these terms
leads to $P(\dethl \vert \dehl) \approx (0.1289)(0.4850)/0.1119 = 0.5586$,
stating that the model shows a 56\,\% probability for finding a {\dethl} that is
within the range of interest. That corresponds to a four-fold increase with respect
to the 13\,\% probability for random picking that corresponds to the 167 species
that meet the criterion in the training set. See the lower limit in Figure
\ref{fig:5}(b) for the full recall with a relevance of zero. For the sake of
completeness, this same analysis can be extended to the opposite case, in which
we measure the ability of the model to discard molecules. We obtained a refusal
probability of $\approx$ 94\,\% that corresponds to an approximate 17-fold
reduction in the number of redundant Kohn-Sham density functional calculations,
as obtained from the complement of this refusal probability,
$1/(6\times 10^{-2})$. These two probabilities show that the graph model
increases the chances for finding materials of interest substantially while
rejecting unsuitable complexes with confidence.

The simpler gradient boosting base-line model results shown in Figure
\ref{fig:5}(b) demonstrate clearly different, inferior performance for the same
testing set. The lower transferability of the base-line model becomes clear,
with an average precision of approximately 40\,\%. In the broader sense, Figure
\ref{fig:5}(b) unveils the effectiveness of using comparatively simple graph
neural networks.

\begin{figure}
  \includegraphics[width=0.5\columnwidth]{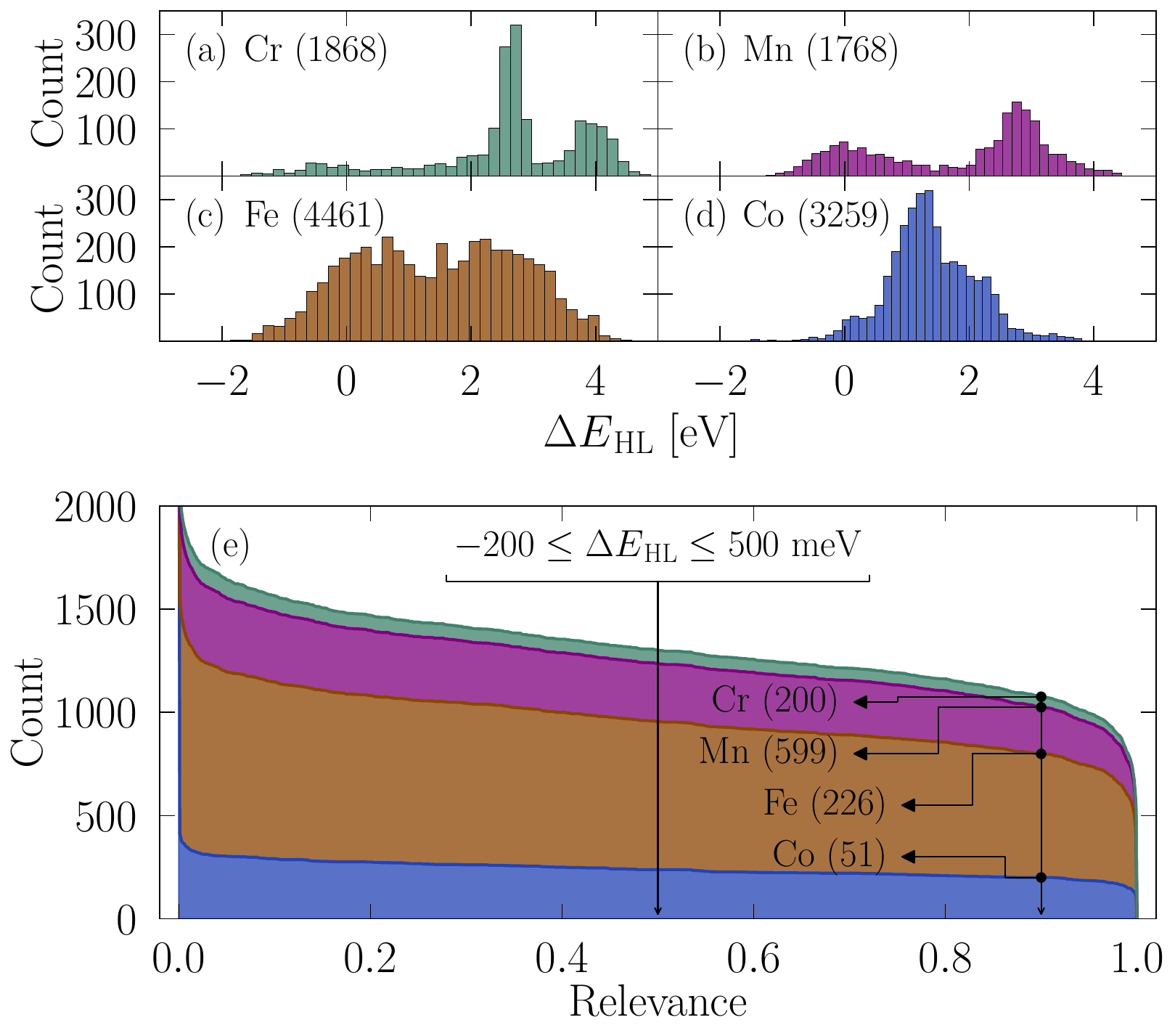}
  \caption{Predicted spin-crossover energies for mononuclear complexes containing
           (a) Cr, (b) Mn, (c) Fe, and (d) Co metal centers, obtained from the
           Cambridge Structural Database, and (e) the total number of plausible
           candidates retrieved for increasingly tightened choices of relevance.
  \label{fig:6}}
\end{figure}
To test our graph neural network system on a much larger variety of mononuclear
transition metal complexes, we collected a total of 11,356 such molecules from
the Cambridge Structural Database spanning the range $13 \leq
N_\mathrm{Atoms} \leq 280$, where $N_\mathrm{Atoms}$ is the number of atoms per
molecule, and used the graph neural network to predict {\dehl}. The data is
included as Supplementary Material. The resulting distributions of the spin-state
energy difference are shown in Figure \ref{fig:6}(a) to (d) for molecules with
Cr, Mn, Fe, and Co centers, respectively. Additionally, in Figure \ref{fig:6}(e)
we report the number of potential spin switching species for the different
metallic centers and extend our evaluation for the number of candidates
retrieved as a function of the relevance criteria.

There are two asymptotic conditions depicted in Figure \ref{fig:6}(e), namely, a
relevance equal to zero that essentially recovers all entries in the data set,
and a relevance equal to one with the opposite outcome. As a reminder, the
boundary for the interval $-200 \leq \dehl \leq 500$ meV by construction is
located halfway through. For that threshold, the classification results in 1,301
species of interest. As stated previously, however, we may increase the
precision of the neural network by tightening the relevance to, e.g., 0.9, and
expect a decreasing recall because the model is not a perfect classifier. As a
result, the final population contains 1,076 metal complexes with the potential
to exhibit spin crossover, with the individual counts reported in Figure
\ref{fig:6}(e). Considering that the test precision shown in Figure \ref{fig:5}
is 80 \% for this value of relevance, approximately 861 of the 1,076 materials
predicted by our model are expected to have a {\dehl} well within the range of
interest, thus vastly accelerating the identification of material candidates.
Interestingly, most of the recovered candidates are Fe-based molecules,
independent of the relevance criteria. This disproportion is somewhat expected
because such 4461 molecules represent the largest population of nearly 39\% in
the data set, as well as the majority with $-1 \leq \dehl \leq 1$ eV. They are
followed in prominence by the Mn-based systems. These observations also bode
well with the known propensity of the Fe(II) complexes to exhibit spin crossover
behavior, as compared to complexes with other relevant metal ions. \cite{M2013,
SPCS2015}

\section{Concluding remarks \label{sec:remarks}}

On the basis of a data set of 1,439 medium-sized transition metal complexes
extracted from the Cambridge Structural Database, we have achieved a
machine-learning model for efficient screening of spin-state-conversion
candidates. The model succeeds by combining descriptors that retain chemically
relevant molecular information with an equivariant graph neural network, and a
subsequent classifier by relevance. Our evaluation shows that use of the neural
network approximately quadruples the chances for finding metal complexes that
might exhibit spin crossover, while confidently rejecting unsuitable molecules.

In this analysis, we assumed the spin-state-conversion energy interval $-200
\leq \dehl \leq 500$ meV to be a reasonable compromise in view of the accuracy
of the exchange-correlation approximation used to equip the data set. The
assumption is of course debatable. The interval of interest depends strongly on
the choice of electronic structure method for the computation of {\dehl} and the
zero-point energy corrections that were not included in our study. Our work,
however, is not focused on addressing refined electronic structure calculations,
but rather on developing high-throughput alternatives to alleviate the computing
burdens required for materials discovery. To this end, our graph model exhibits
a competitive ability to exclude unsuitable materials that, in turn, results in
approximately a 17-fold reduction of low-productivity computations.

\section*{Supplementary Material \label{sec:material}}

The Supplementary Material includes comma separated value files for the two
datasets used in this work. These include the CSD identifier, {\dehl}, and the
transition metal center for every species listed therein.
  
\begin{acknowledgement}
  This work was supported as part of the Center for Molecular Magnetic Quantum
Materials, an Energy Frontier Research Center funded by the U.S. Department of
Energy, Office of Science, Basic Energy Sciences under Award No.
\mbox{DE-SC0019330}. This research used resources of the National Energy
Research Scientific Computing Center (NERSC), a Department of Energy
Office of Science User Facility using NERSC award \mbox{BES-ERCAP0022828}.
\end{acknowledgement}

\bibliography{references}

\end{document}